ARTICLE

# Freedericskz Transitions in the Nematic and Smectic $Z_A$ Phases of DIO

J.T. Gleeson[a]; S.N. Sprunt,[a]; A. Jákli,[a,b] P. Guragain,[c] R.J. Twieg[c]

The remarkable material DIO presents fascinating behaviors. It has been extensively studied as one of the first materials exhibiting a ferroelectric nematic phase. However, at higher temperatures it exhibits what has been termed the Smectic $Z_A$: identified as an orientationally ordered, antiferroelectric phase with a density modulation in direction perpendicular to the optic axis. At even higher temperatures, this transitions to an apparently normal nematic phase. We have studied the splay-bend Freedericksz transition in the nematic and $SmZ_A$ phases of the material DIO. Both the magnetic and electric field transitions were utilized. We observed the transitions by measuring effective birefringence and capacitance as well as with polarizing light microscopy. In both the nematic and the $SmZ_A$ states the field induced transitions resemble (in numerous aspects) the classical Freedericksz transition. These enable determinations of several fundamental material parameters and also reveal intriguing aspects of the $SmZ_A$ phase, including the surprising behavior of the elastic constants and the dielectric anisotropy. Detailed comparison with Frank elastic theory of the Freedericksz transition [1] shows that the N phase behaves largely as expected, but the transition in the $SmZ_A$ phase differs significantly. Two specific examples of this are the onset of striations in the Freedericksz distorted state, and the presence of optical biaxiality. The former may be related to the periodic Freedericksz transition [2] as it coincides with a large increase in the splay elastic constant. The latter has been predicted for the $SmZ_A$ phase, but not previously observed. [3]

## Introduction

The behavior of nematic liquid crystals (NLC) under external electric and/or magnetic fields is extremely well studied. Indeed, using an electric field to control the optical axis of a NLC underpins the overwhelming majority of NLC technologies. This enormous body of work provides a useful framework with which we can place the new liquid crystal mesophases, such as the Smectic $Z_A$ ($SmZ_A$) studied here into context. [4] More specifically, the Freedericksz transition (FT), [5,6] in which the orienting effects of confining substrates are in opposition to the effect of an external field, has been exhaustively examined, both theoretically and experimentally. The “classical” FT is characterized by a non-zero threshold field at which the optic axis begins to rotate away from its zero-field state. Upon extrapolating the field to infinite strength, the optic axis becomes parallel to the field direction.

a. Department of Physics, Kent State University, Kent, OH 44242, USA
b. Materials Science Graduate Program and Advanced Materials and Liquid Crystal Institute, Kent State University, Kent, OH 44242, USA
c. Department of Chemistry and Biochemistry, Kent State University, Kent, OH 44242, USA

This work presents a detailed study of both the electric and magnetic field induced FT in the of the material DIO. [7] This material, in addition to exhibiting the polar, ferroelectric nematic ($N_F$) phase (which is not the topic of this work) also exhibits both the uniaxial, apolar nematic (N) phases and the newly described Smectic $Z_A$ ($SmZ_A$) phase. [3] Both the N and $SmZ_A$ exhibit both electric and magnetic field induced FT’s. All transitions are qualitatively similar, although the differences are both important and illuminating. Tracking how these transitions evolve as the material progresses between the two phases brings this into even sharper focus.

In particular, we focus on the splay-bend FT [5], in which a LC is confined between parallel plates separated by a distance, *d*. The plates are treated so as to induce uniform surface alignment (defined as the *x* direction). Because of the anisotropy in either the dielectric tensor or the diamagnetic susceptibility tensor, an external field exerts a torque on the LC director. When the applied field exceeds its threshold, the director rotates in the *z* direction. For the magnetic case, the threshold field is $\boldsymbol{B_c} = \frac{\pi}{d}\sqrt{\frac{K_{11}\mu_0}{\Delta\chi}}$, where $K_{11}$ is the splay elastic constant, $\mu_0$ is the permittivity of free space, $\Delta\chi$ is the anisotropy in the diamagnetic susceptibility. Thus, studying this transition yields important information on the balance between elastic and external field effects. There is copious literature on every aspect of this effect. [1,8]

How will the $SmZ_A$ phase be affected by external fields? Two attributes of the $SmZ_A$ phase can be expected to be significant. In addition to the elastic energy modes associated with distorting the nematic director, we can anticipate there will be additional modes involving distortion of the layered structure. Moreover, as is common in all smectic phases, modes which would require layer compression or dilation should be forbidden. That is, any director mode that requires rotation out of the layer plane, or which would necessitate layer rupture is therefore strongly suppressed. At present, there is no comprehensive description of the elastic energy of the $SmZ_A$ phase. Furthermore, the antiferroelectric nature of the $SmZ_A$ phase in DIO cannot be neglected. While antiferroelectric materials have no net polarization at zero applied electric field, they should be substantially more polarizable than the nematic phase. Because of this, we might anticipate the coupling to an applied electric field to be rather more complex than the well-known director torque via the anisotropy in the dielectric constants.

## Experiments

The material DIO [9](2,3′,4′,5′-tetrafluoro [1,1′-biphenyl]-4-yl 2,6-difluoro-4-(5-propyl-1,3-dioxan-2-yl) benzoate was resynthesized [10] as described in SI. Its purity was confirmed using NMR and DSC; moreover, all phase transitions as described below were sharp, well-defined and in agreement with published values. [11] This compound has two stereo isomers, and at elevated temperatures, isomerization may take place. [12] As described in the SI, we used isomerically pure (*trans*) DIO, and took care to never heat it about 100°C. Using this material, we filled sandwich cells having rubbed polyimide surface treatment to create planar alignment. Before filling, the plate separation, *d*, was measured interferometrically, and the capacitance of the empty cell recorded. Cells were filled using capillary action at 100°C on a hotplate We observed uniform planar alignment in both the N and $SmZ_A$ phases of DIO.

With planar alignment, the splay-bend Freedericksz transition (FT) [5] is possible using an external field perpendicular to the alignment direction (and hence the substrates). We explored this using both ac electric field (at various frequencies) and dc magnetic field. Furthermore, as this FT corresponds to an increase of the angle between the director and the planar aligning surfaces ($\theta$), we employed three complementary probes to detect and monitor this director distortion: capacitance, optical phase difference and microscopy.

For all measurements, the samples were held in a temperature-controlled enclosure. As previously noted, we took pains that the sample temperature never exceeded 100°C. The capacitance (and loss) of the material was determined by applying a sinusoidal voltage of varying frequency and amplitude between the ITO electrodes and measuring the in-phase and out-of-phase current. For this we used a variable gain current-to-voltage pre-amplifier. Before making any measurements, the NLC cell was substituted for purely resistive load and the lock-in amplifier phase angle adjusted so that the out-of-phase current registered at zero. This step eliminates the influence of residual capacitance in lead wires, etc. The entire circuit was then tested using a standard capacitor. With this arrangement we can measure the capacitance and loss of the LC sample subject to a sinusoidal potential difference having amplitude varying between $0.004V_{rms}$ and $5V_{rms}$, and with frequency between 10Hz and 50kHz.

In order to study the magnetic field induced Freedericksz transition, the LC sample was suspended in the bore of a 10T superconducting solenoid. For these experiments, the magnetic field direction was perpendicular to the glass plates (and the alignment direction. With this we can measure the capacitance (and loss) of the LC layer at various magnetic fields. Furthermore, the housing had a 2mm through hole for optical access. By positioning a lamp and a polarizer inside the magnet, and with use of a long-distance, reflecting microscope we could obtain polarizing optical microscopy images of the LC layer subject to magnetic fields. However, the constraints of the magnet and the vibrations of the cooling system limited both the magnification and resolution.

In addition to the above, we also performed optical phase difference (OPD) measurements for the electric-field induced Freedericksz transition. For these we employed the standard photo-elastic modulator technique, [13] with additional phase compensation to establish the zero OPD configuration.

## Results

In order to understand the baseline of field induced effects in this material, we begin with a comprehensive investigation of the nematic phase in DIO. Prior to any measurements, the textures of the samples were examined using polarized optical microscopy (POM). In all cases, uniform, planar alignment was observed throughout both the N and the $SmZ_A$ phases. Indeed, the transition from N to $SmZ_A$ is rather subtle as observed in the microscope, as has been previously reported. [3]

A sandwich cell with planar alignment corresponds to the nematic director perpendicular to the substrate normal (with no applied field). Before filling with DIO, the capacitance of the empty cell, $\boldsymbol{C_0} = \frac{\boldsymbol{\varepsilon_0 A}}{\boldsymbol{d}}$, where *A* is the electrode area, *d* is electrode separation was measured. Therefore, the measured capacitance is given by $\boldsymbol{C} = \frac{\boldsymbol{\varepsilon_\perp \varepsilon_0 A}}{\boldsymbol{d}}$ when $V_0$, the potential difference used to make the measurements is small. The ratio $C/C_0$ then yields $\boldsymbol{\varepsilon_\perp}$ Figure 1 shows $\boldsymbol{\varepsilon_\perp}$ vs temperature at a variety of different frequencies. Both the N-$SmZ_A$ and the $SmZ_A$ -$N_F$ transitions are prominent. The values shown for the $N_F$ phase were based upon the capacitance as actually measured. We note that there are unresolved issues concerning this; [14] however these are not within the scope of the present paper. We do not represent that values shown here below the $SmZ_A$ -$N_F$ transition temperature are reflective of the $N_F$ state.

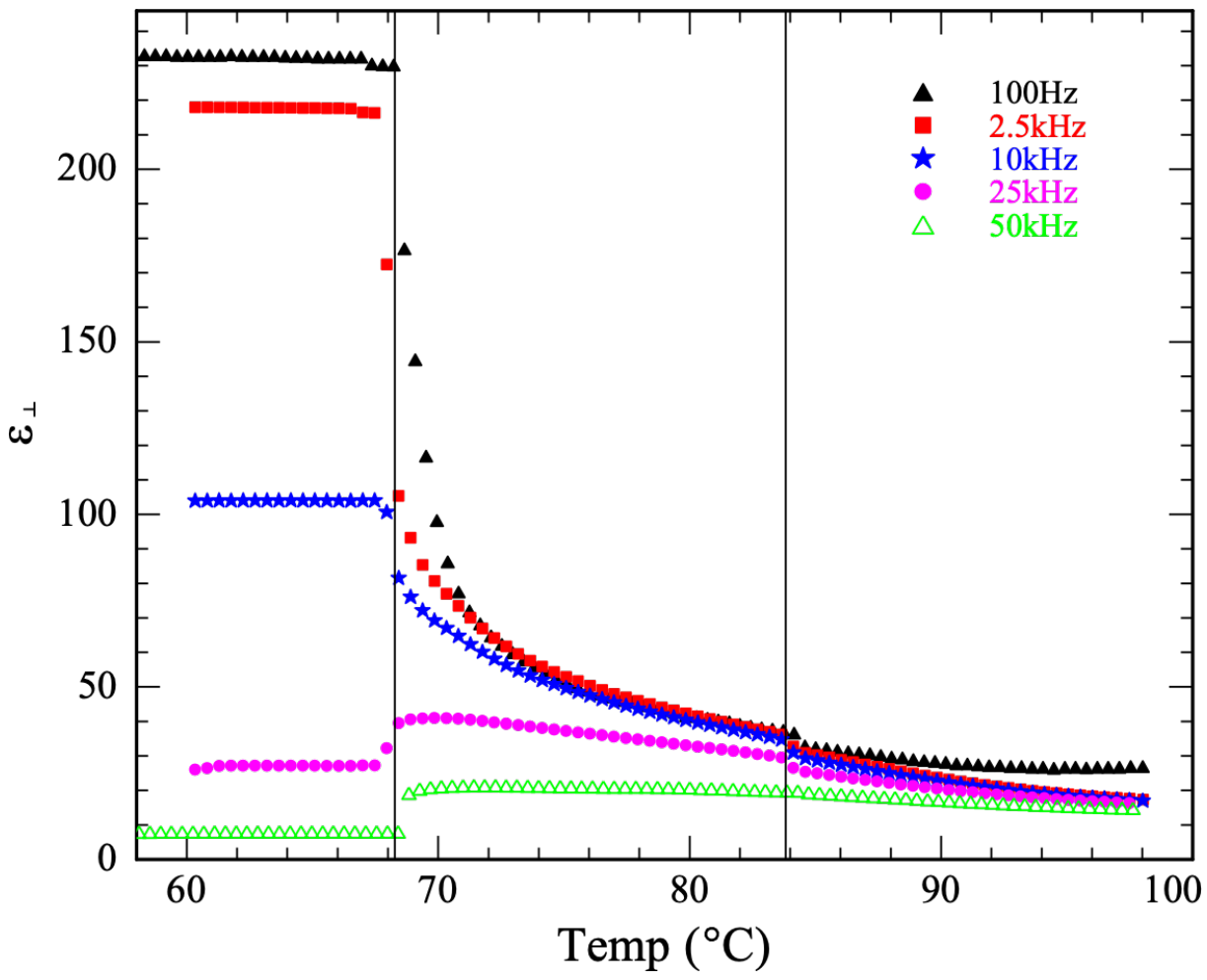

Figure 1 Perpendicular component of dielectric constant vs temperature at various frequencies. The vertical lines indicate the N- $SmZ_A$ and the $SmZ_A$ -NF phase transitions.

We examined both the magnetic field and electric field FT in the nematic phase of DIO. Note that these studies do not span the entire nematic phase range, as we took cautions to remain below 100°C as described above. Both effects were studied via measurement of capacitance, and the electric-field induced FT was also studied via OPD measurement. We also obtained POM images of the texture during both transitions. In the nematic phase, we observe the classic Freedericksz transition. At voltages well below and well above the transition we observe a uniform texture. Frequently, as the voltage is increased above the threshold we temporarily observe walls. [15] These disappear within a few seconds. The electric field induced transition was studied at frequencies ranging between 100Hz-50kHz. Figure 3 shows an example of how the effective birefringence, $\boldsymbol{n_{eff}} - \boldsymbol{n_0}$ depends on applied voltage.

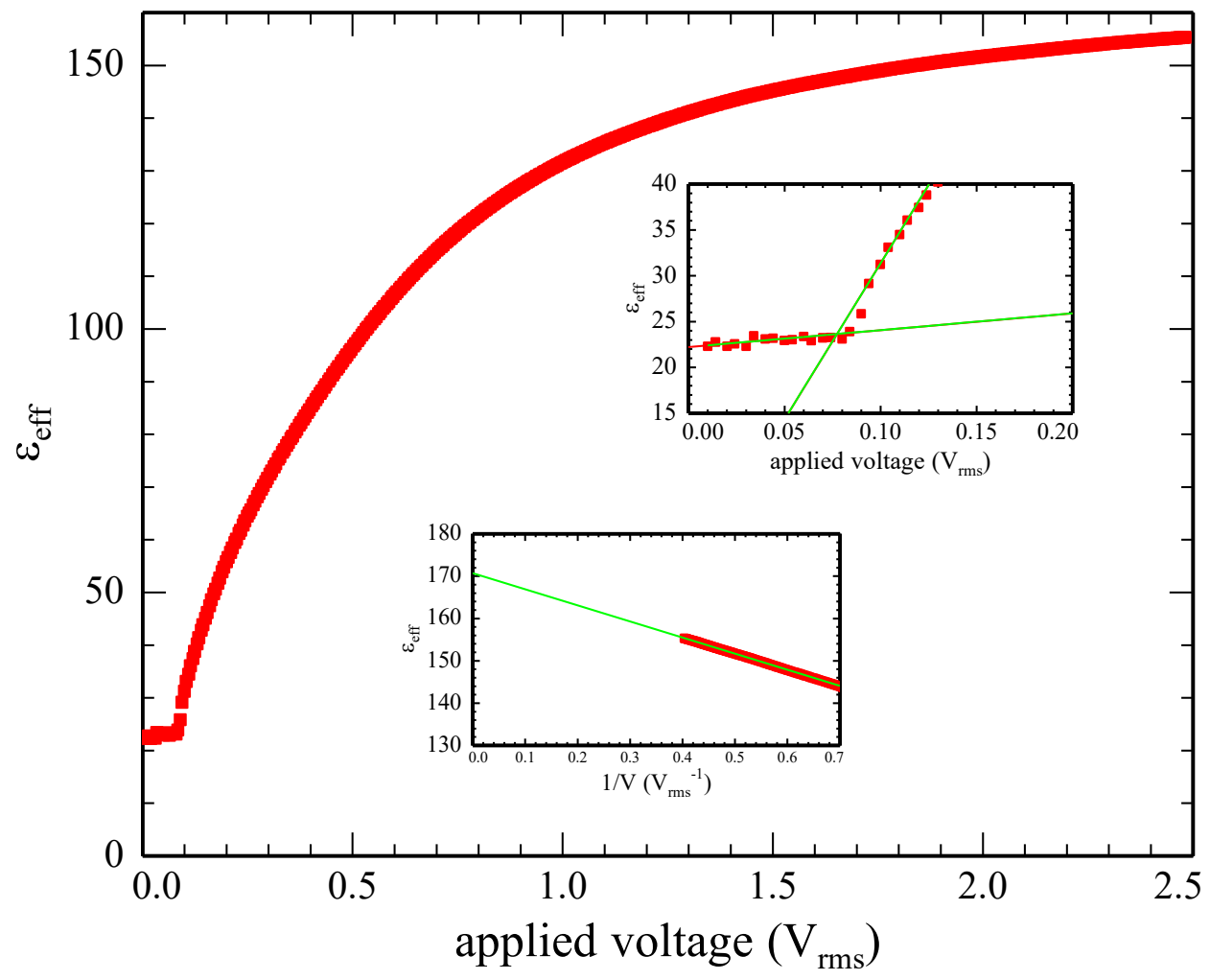

Figure 2 Effective dielectric constant vs voltage at 90°C and 2.5kHz. The upper inset shows the transition occurring at 0.076Vrms. The lower inset shows the extrapolation to infinite voltage enabling the determination of $\varepsilon_{\parallel}$.

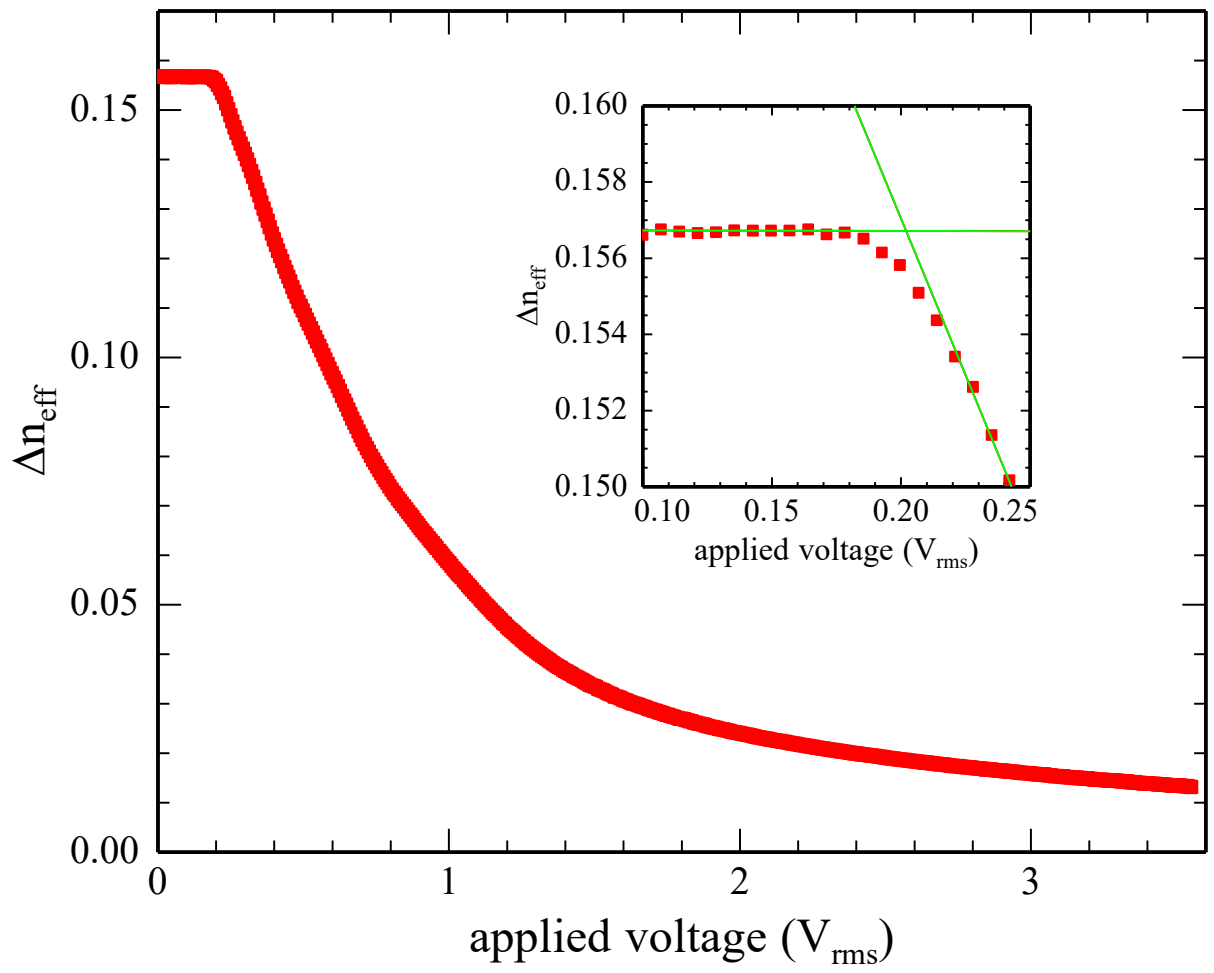

Figure 3 Effective birefringence vs applied voltage for DIO at 84°C and 2.5kHz. Inset shows close up of Freedericksz transition along with the fitting used to determine the threshold voltage.

Capacitance measurements are particularly valuable in studying the Freedericksz transition. The transition may be induced by the probe field, which can be of variable frequency. Moreover, the values measured both at very small external field (i.e. below the threshold) and, by extrapolating to infinite field reveal the perpendicular and parallel components of the dielectric tensor ($\boldsymbol{\varepsilon}_{\perp}$ and $\boldsymbol{\varepsilon}_{\parallel}$) , (at the desired frequency). [16] The threshold FT voltage, given by $\boldsymbol{V_c} = \boldsymbol{\pi}\sqrt{\frac{\boldsymbol{K_{11}}}{\boldsymbol{\varepsilon_0 \Delta\varepsilon}}}$, which is straightforward to measure, yields the splay elastic constant. An example of capacitance measurements during the electric field induced FT is shown in Figure 2. Lastly, an example of the capacitance during the magnetic field induced FT is shown in Figure 4. In the nematic phase, both types of FT are useful for determining material parameters as described below. Specifically, if we plot these measurements in a reduced fashion, i.e. plotting $\frac{\boldsymbol{\epsilon_{eff}}}{\boldsymbol{\epsilon_{\parallel}}} - \boldsymbol{1}$ vs $\frac{\boldsymbol{B}}{\boldsymbol{B_c}} - \boldsymbol{1}$ (or $\frac{\boldsymbol{V}}{\boldsymbol{V_c}} - \boldsymbol{1}$) , elastic theory predicts the curve obtained should be linear right above the transition, having slope $\frac{\boldsymbol{2\gamma}}{\boldsymbol{1+\kappa}}$ (for the magnetic FT), where $\boldsymbol{\gamma} = \frac{\boldsymbol{\epsilon_{\parallel}}}{\boldsymbol{\epsilon_{\perp}}} - \boldsymbol{1}$ and $\boldsymbol{\kappa} = \frac{\boldsymbol{K_{33}}}{\boldsymbol{K_{11}}} - \boldsymbol{1}$. For the electric case, the slope will be $\frac{\boldsymbol{2\gamma}}{\boldsymbol{1+\kappa+\gamma}}$. [16] Thus, with these two measurements we can obtain the dielectric anisotropy, the splay and bend elastic constants, and the diamagnetic anisotropy. The temperature dependence of both elastic constants is shown in Figure 5. These results are comparable with Ref. [12] where measurements solely in the N phase are reported.

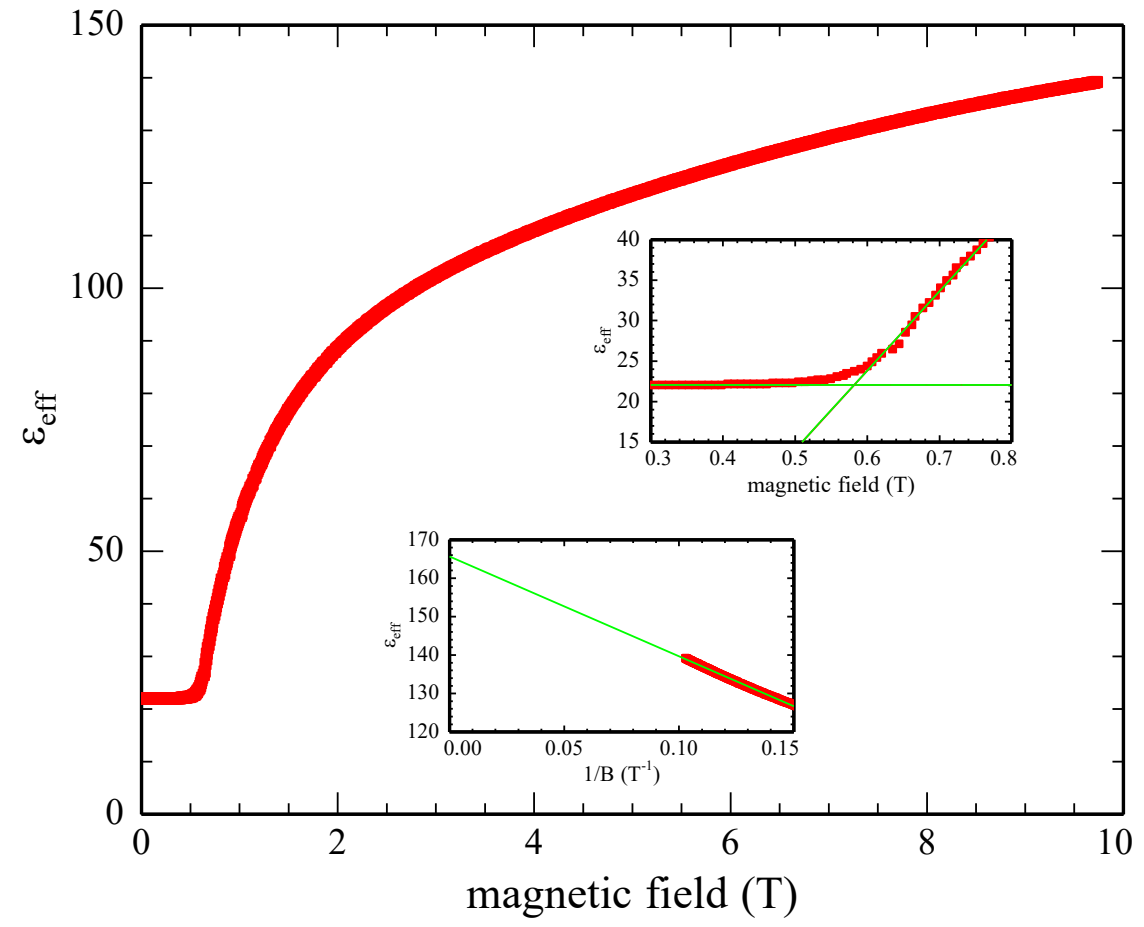

Figure 4 Effective dielectric constant vs magnetic field at 90°C. The probe voltage is $0.00V_{rms}$ at 2500 Hz. Upper inset shows close-up of FT; lower inset shows extrapolation to infinite magnetic field.

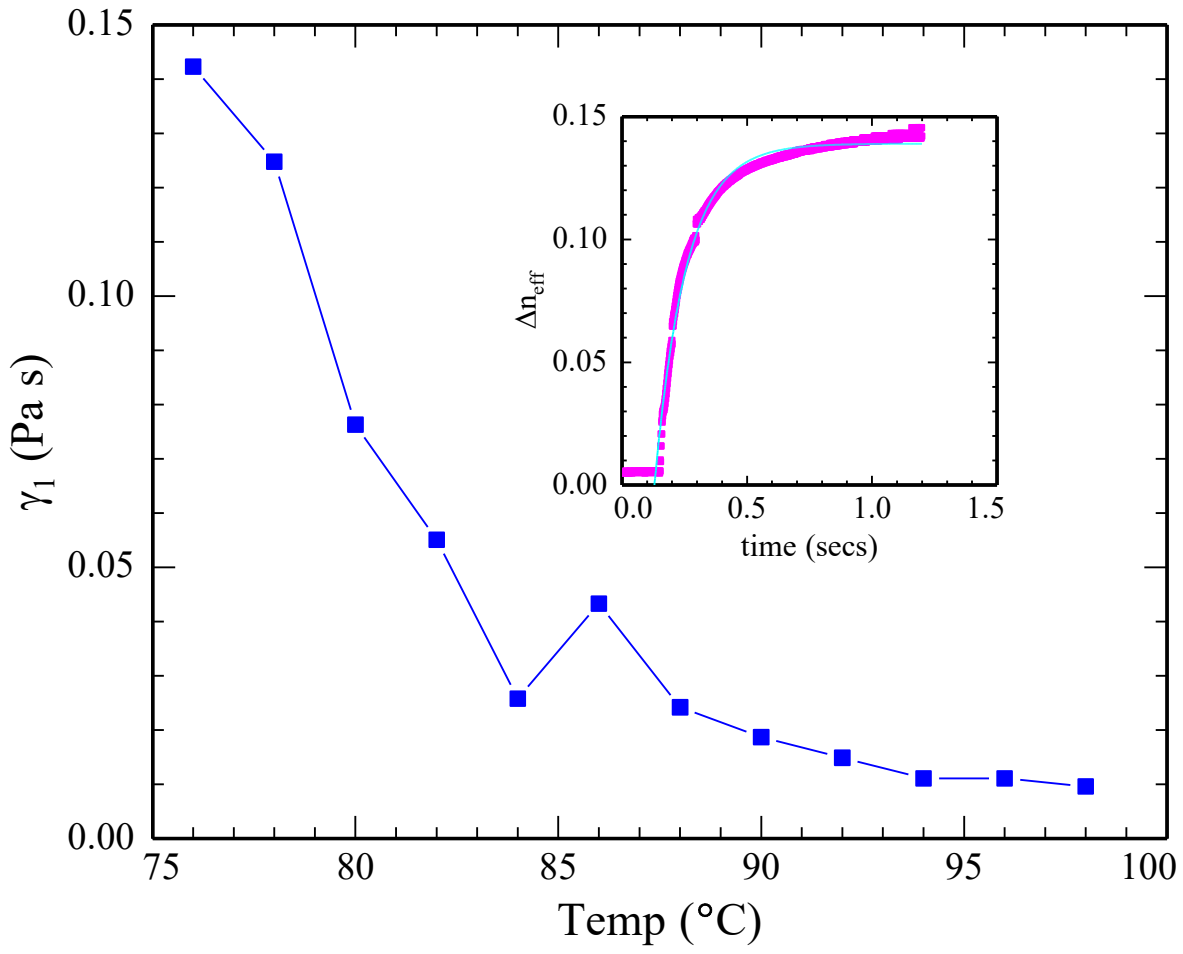

Figure 6 Rotational viscosity vs temperature. Inset: example of change in effective birefringence vs time after stepwise decrease in applied voltage. Green line shows fit to single exponential.

The dynamics of the FT can yield some insight into LC flow properties. Specifically, upon stepwise reduction of the external field (from above threshold to zero), the effective birefringence decays exponentially with time constant $\tau_0 = \frac{\gamma_1 d^2}{K_{11}\pi^2}$, where $\gamma_1$ is the orientational viscosity. Although backflow effects are expected in this geometry, their presence should not affect the time constant. [17] An example of the birefringence decay is shown in Figure 6 ; the inset shows the dependence of $\gamma_1$ on temperature. It is notable that while the viscosity increases as you enter the $SmZ_A$ from the N phase, there is no abrupt change. Furthermore, and the values of $\gamma_1$are not significantly different from calamitic nematic materials.

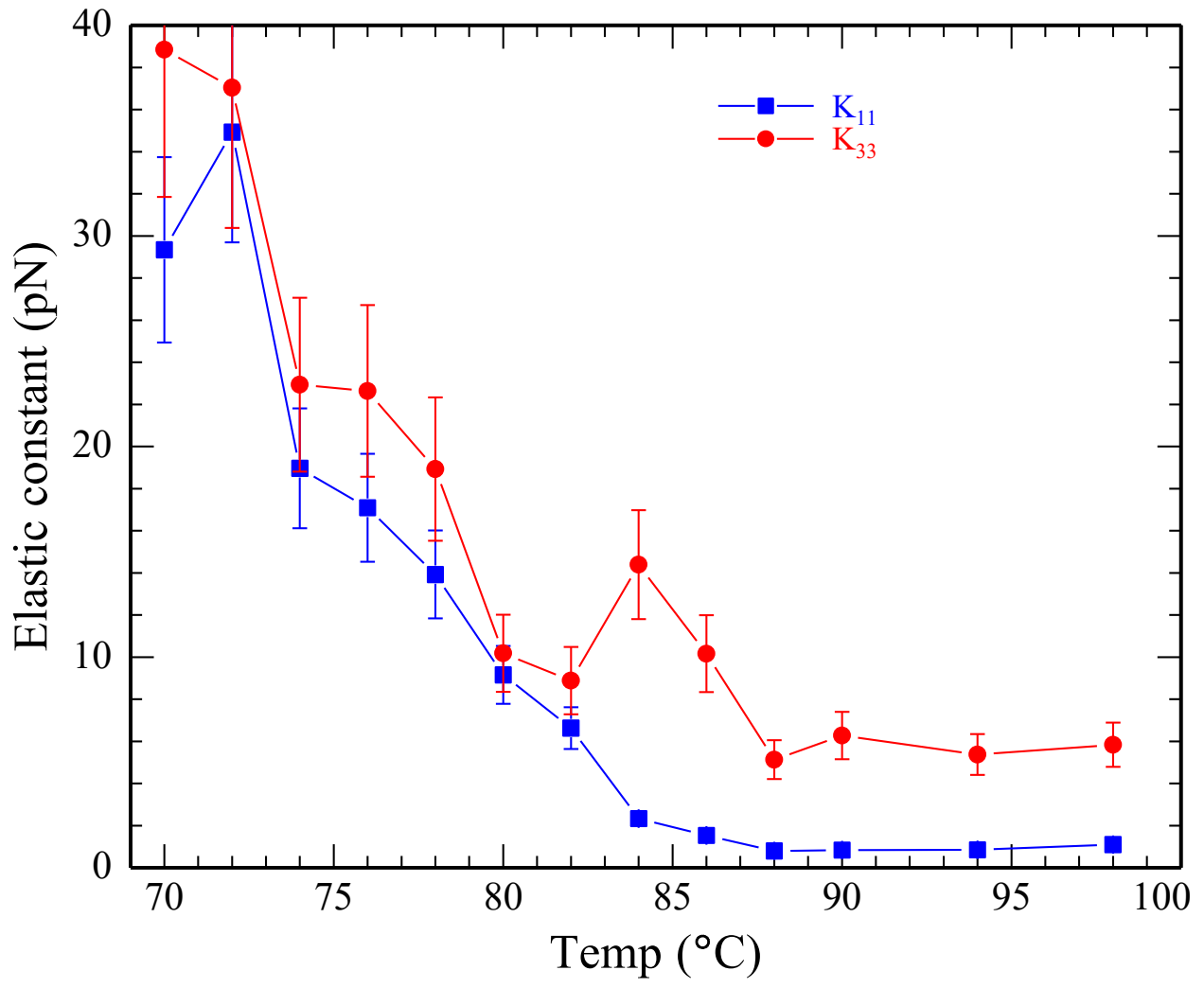

Figure 5 Splay and bend elastic constants of DIO vs temperature. Both increase dramatically at the N-$SmZ_A$ transition. The error bars reflect the uncertainty in determining both the threshold potential difference and the slope at onset (for K33) $SmZ_A$

Reference [3] posits that aligned samples of the $SmZ_A$ phase will be biaxial, there will be some residual birefringence when the principal axis of the refractive index tensor is parallel to the optical wavevector. This is not surprising in that there are two distinct directions in the plane perpendicular to the director as this plane contains the layer normal. Nonetheless optical biaxiality in liquid crystals is typically difficult to measure. [18,19] Our approach uses an external field to unambiguously align the principal optic axis and then test for residual birefringence in the plane perpendicular to it. [20–22] Uniaxial materials should extrapolate to zero residual birefringence at infinite field, as in this case the single optic axis will be collinear with the field. However, this will only be strictly true in the case of infinite surface anchoring potential, which is certainly not the case in practice. In the case of finite surface anchoring, [23] the residual birefringence will reach zero at finite applied field, and so the extrapolated value will be negative; we can see this above the N-$SmZ_A$ transition in Figure 7. This figure clearly shows the onset and increase of biaxiality as one cools below the N- $SmZ_A$ transition temperature.

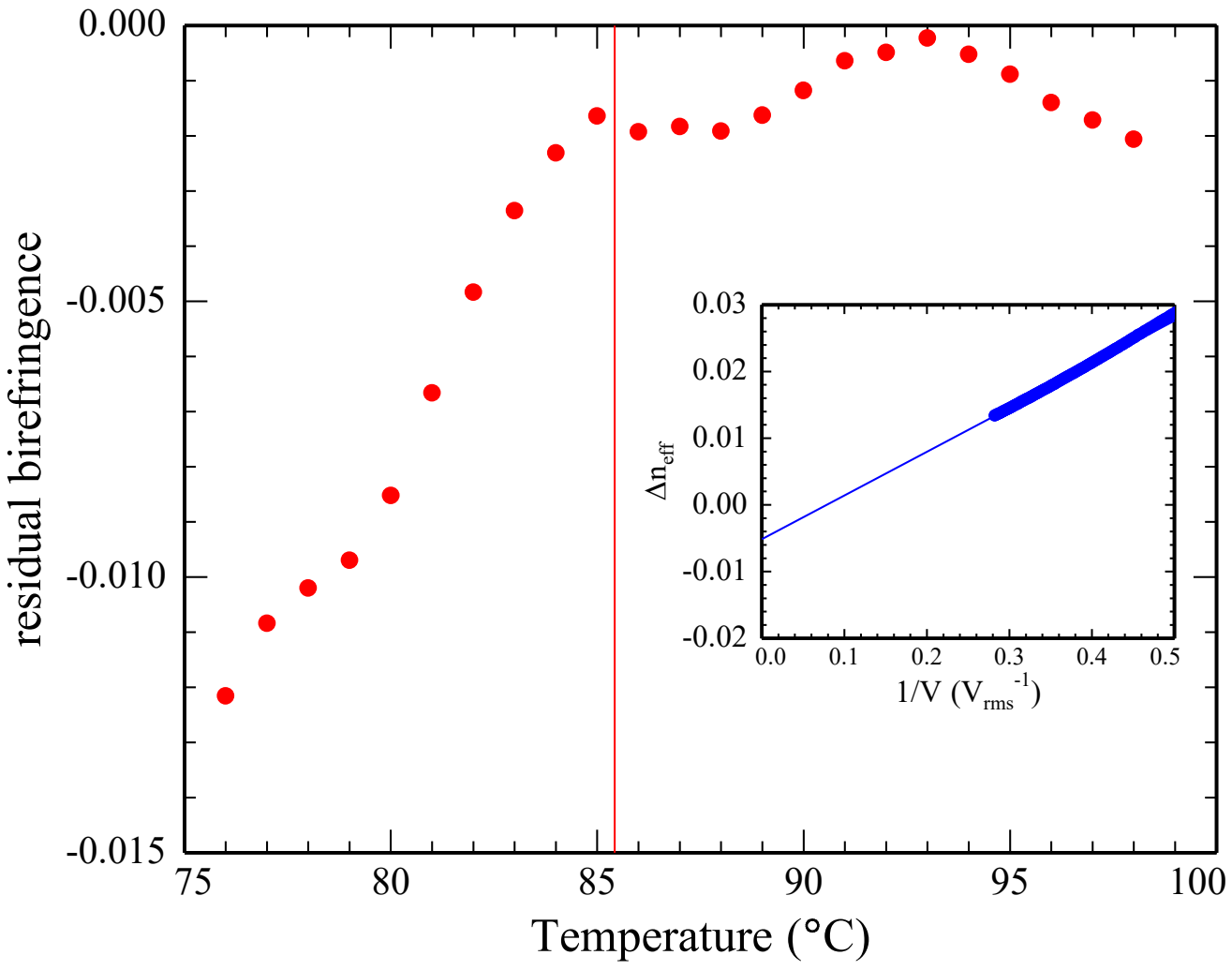

Figure 7 Effective birefringence extrapolated to infinite potential difference vs temperature. Vertical line is the N-SmZa transition. Inset: extrapolation example (corresponds to 80°C)

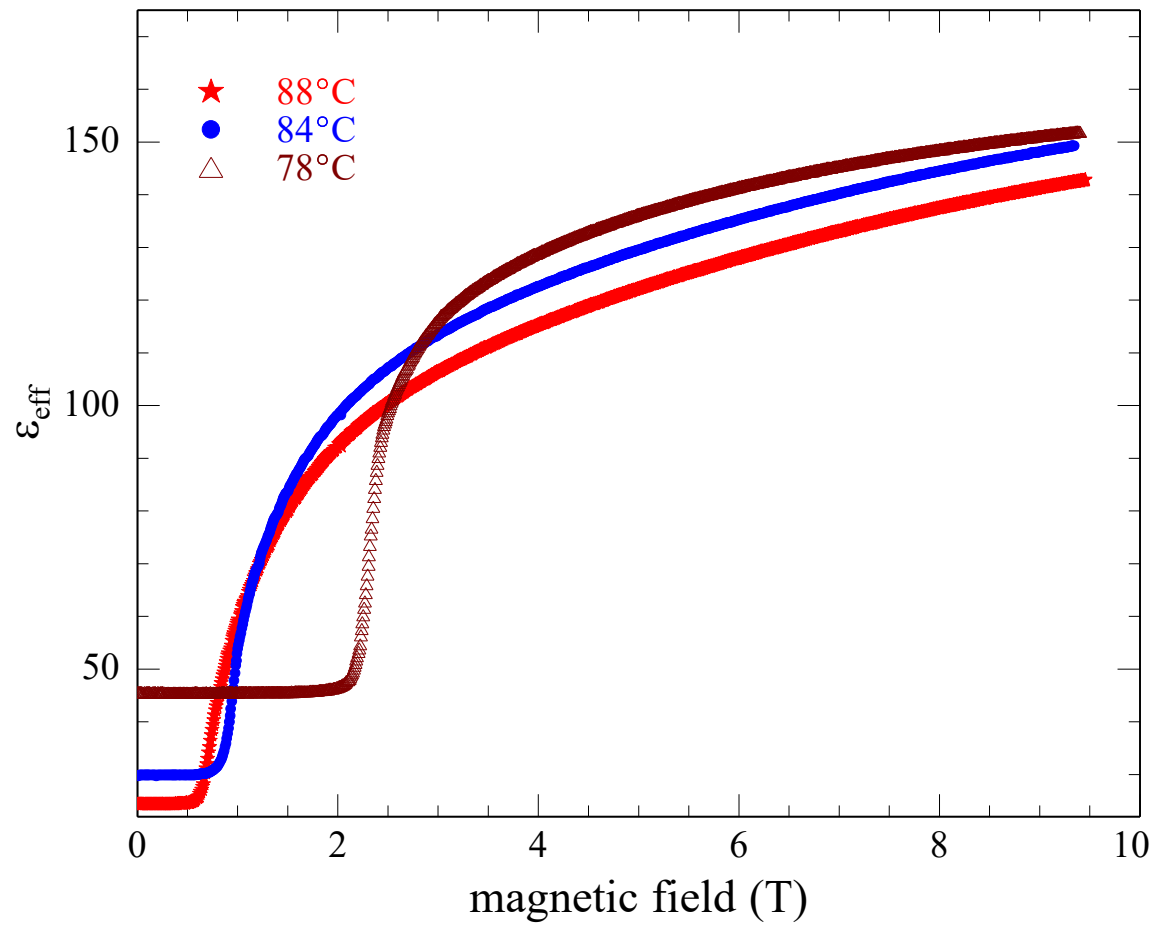

Figure 8 Magnetic field induced FT at different temperatures above and below the N-SmZa transition. No significant increase in rounding is seen below the transition temperature.

## Discussion

The results above show that both (electric and magnetic) field induced splay-bend Freedericksz transitions occur in N and $SmZ_A$ phases of DIO. The observations in the nematic phase are of course to be expected. In the the $SmZ_A$ phase it is less clear what is to be expected. It is instructive to look more closely at this transition, and to examine in detail how the both the transition and the distorted state above it differ as the material enters the $SmZ_A$ phase from the nematic.

In the ideal FT, (i.e. infinite anchoring strength, zero pre-tilt angle), at the threshold field the measured quantity (capacitance or birefringence) will exhibit a discontinuity in its slope. When these assumptions are relaxed, the transition will become rounded. [24,25] In DIO, this is also observed, but the rounding becomes far more pronounced in the $SmZ_A$ phase, as is shown in Figure 9. While it is difficult to ascertain the exact cause for the enhanced rounding, it would seem unlikely that the onset of the $SmZ_A$ is accompanied by a significant increase in pre-tilt angle. Moreover, were this the case, similar behavior would be seen in the magnetic FT. Figure 8 shows that this is not the case. In other words, the origin of the dramatic rounding of the electrically induced transition is unclear. This effect also contributes to the uncertainty in determining the elastic constants, as is reflected in the error bars in Figure 5.

The FT in DIO, especially in the $SmZ_A$ phase, exhibits further discrepancies from the ideal, nematic FT. The standard Oseen-Frank theory for the FT predicts that at the threshold field there is an exchange of stability between the undistorted and the distorted states. That is, right above the threshold the undistorted state is unstable, and right below the distorted state becomes unstable. Therefore, in the well-known nematic FT one expects no difference no matter whether the external field is increasing or decreasing, except perhaps if the field is changed more rapidly than the LC's ability to respond. This is indeed what is observed with DIO in the N phase, for electric or magnetic fields. In the $SmZ_A$ phase, we observe rather different behavior. An example of the magnetic FT is shown in Figure 10. In this case we observe a 30% increase in the critical field depending on whether it is increasing or decreasing. The overall curve is not substantially altered. In other words, the elastic resistance to the director distortion is lower (i.e. lower $K_{11}$) when the field is increased (from below to above threshold) than when the field is decreased. The electric FT in the $SmZ_A$ phase is even more intriguing (see Figure 9). In this case, the $SmZ_A$ phase not only shows a difference in threshold voltage between increasing and decreasing, but the overall shape of the response is dramatically different. The transition is more rounded, and the initial slope is very different.

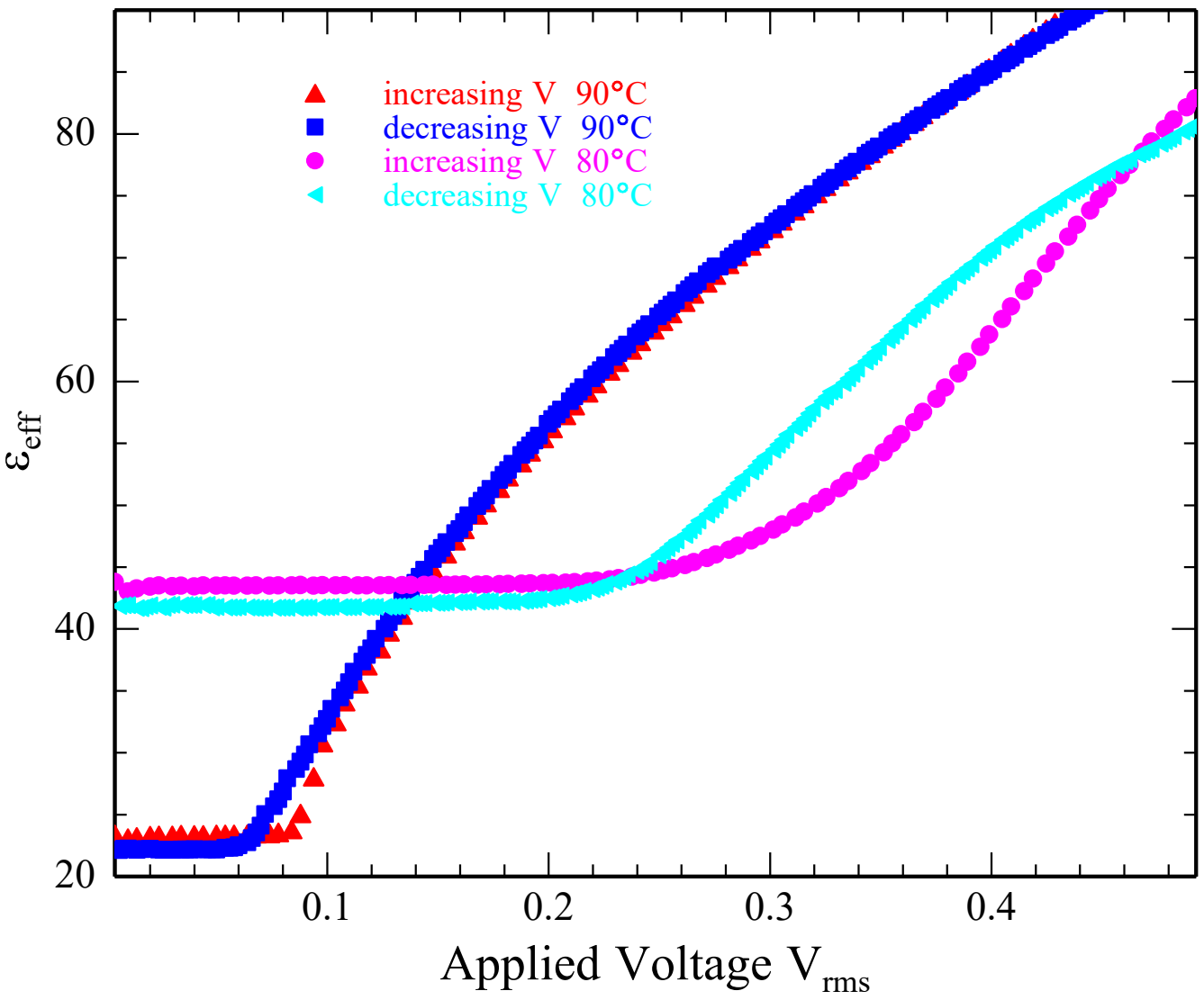

Figure 9 Effective dielectric constants for both increasing and decreasing applied voltage in both the N and SmZa phase.

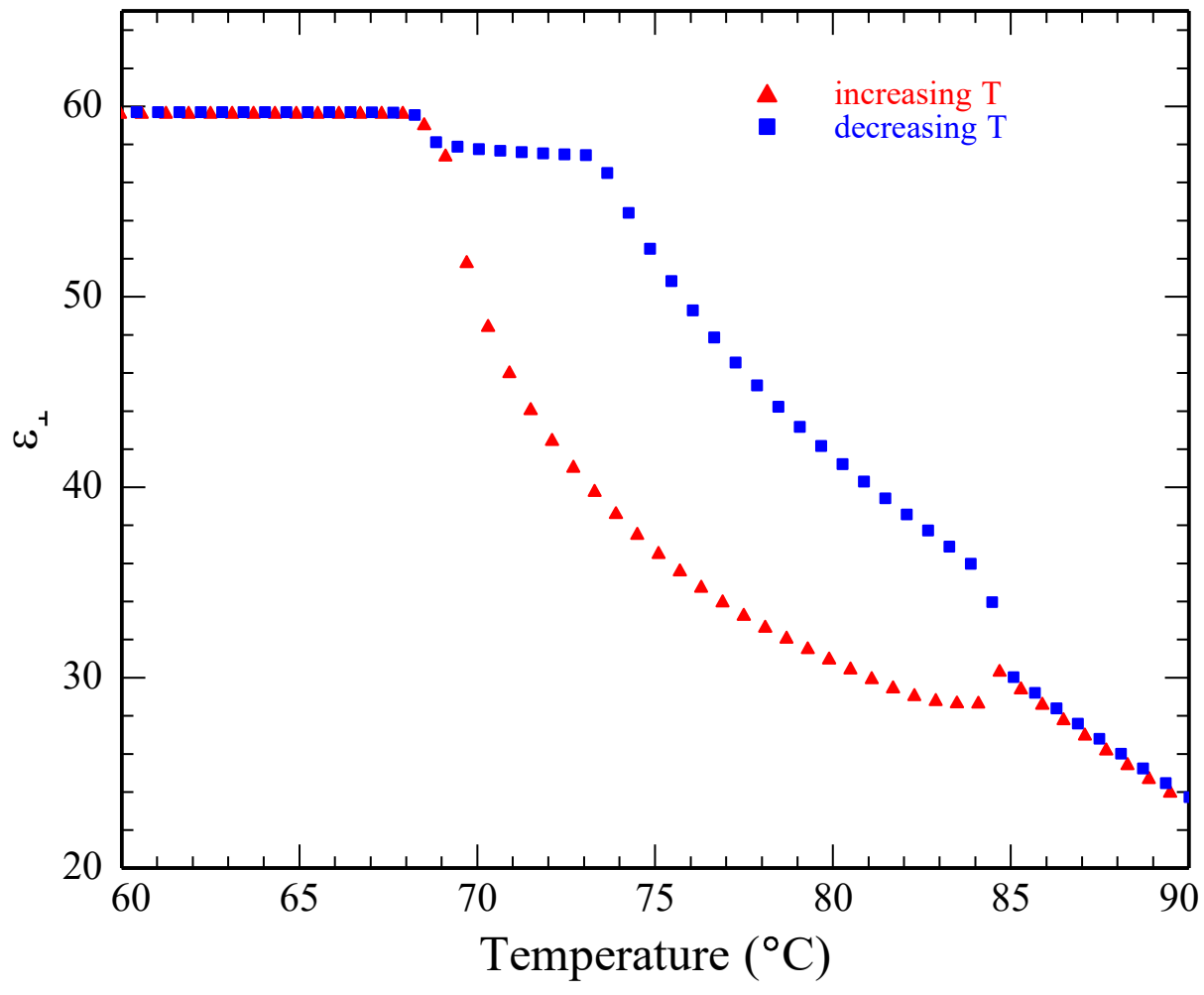

Figure 11 Temperature dependence of $\varepsilon_{\perp}$ during both heating and cooling. Measured at 10kHz and probe voltage of 0.05$V_{rms.}$ Note the unusual kink at 73.3C, but only during cooling.

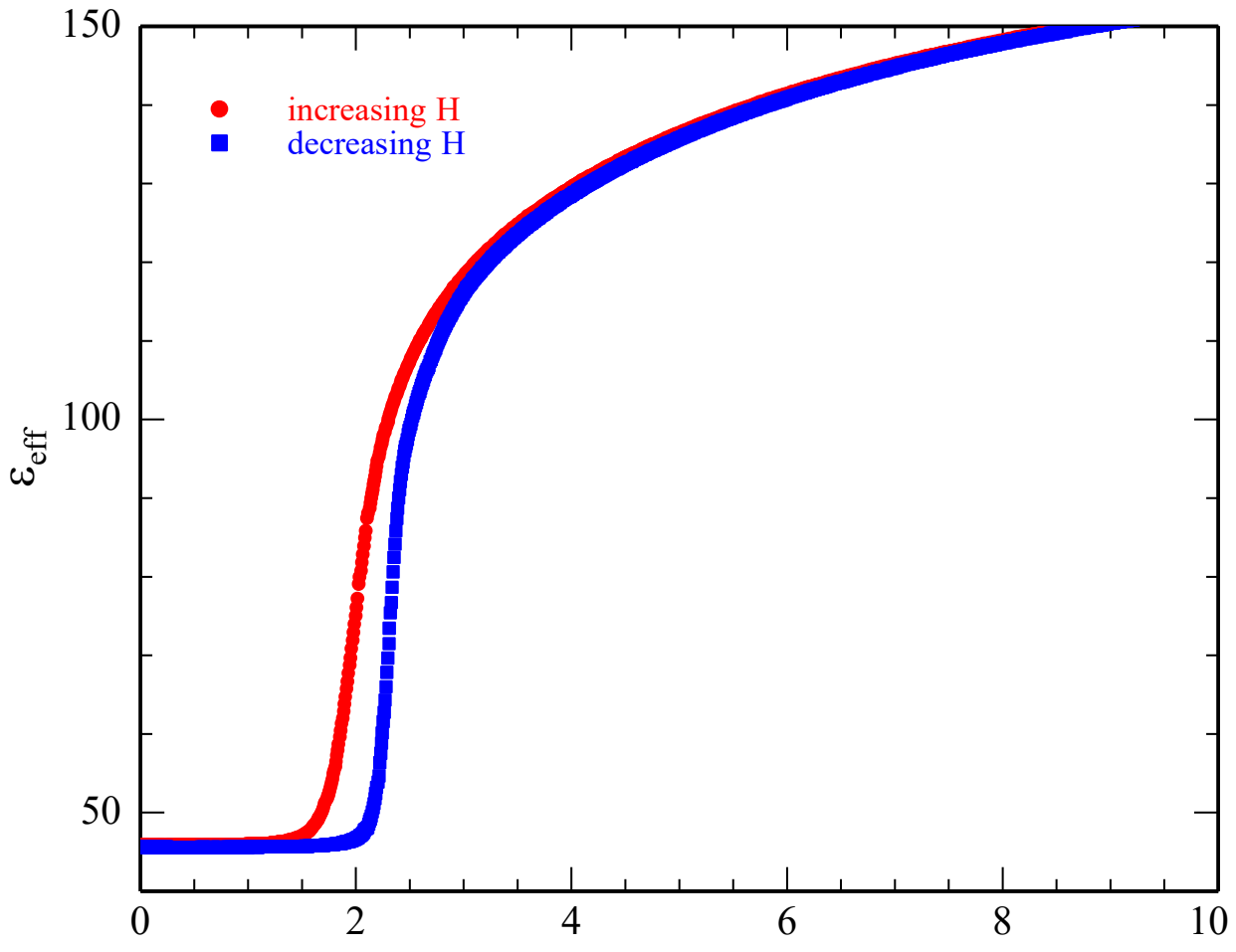

Figure 10 Hysteresis of the magnetic FT in SmZa phase of DIO. Temperature 78°C; probe voltage 0.004$V_{rms}$; 2.5kHz

Indeed, the FT is not the only circumstance where hysteresis is observed, although it is the most prominent; Figure 1 shows the dependence of $\boldsymbol{\varepsilon}_{\perp}$ on temperature. When this is examined more closely, comparing cooling from above the N- $SmZ_A$ phase transitions to and heating from below the $N_F$- $SmZ_A$ transition, $N_F$ phase, a small but significant difference emerges, but only within the $SmZ_A$ phase – c.f. Figure 11. We further note that no difference is discernible via POM comparing heating and cooling through the $SmZ_A$ range. Possible origins of these hysteretic behaviors are discussed subsequently.

What is the origin of these hysteretic behaviors? In both the N and $N_F$ phases, there is no hysteresis between heating and cooling; in the $SmZ_A$ phase it is unmistakable and repeatable. In the N phase, during both electric and magnetic field-induced FT's, there are only minor differences between increasing and decreasing applied fields. These differences are common in traditional nematics, and usually disappear when the fields' rate-of-change is sufficiently low. This is not the case in the $SmZ_A$ phase. In the magnetic FT, there is a persistent difference in the critical field between increasing and decreasing H. In the electric FT, not only is the critical voltage hysteretic, but there is also a distinct difference in the observed shape of the curve depending on direction.

Figure 11 probes the component of the dielectric constant perpendicular to the nematic director, which is dictated by the planar surface alignment treatment on the glass plates. Unspecified is the direction of the smectic layering in the $SmZ_A$ state. Two extremes are imagined in Ref. [3] : in one the layer normal (which must be perpendicular to the director) lies in the plane of the glass plates (the "bookshelf" geometry); in the other the layer is perpendicular to the glass plates (the "parallel" geometry). Other than by x-ray scattering techniques, (which are not practical in thin film geometries), unambiguous determination of the layer normal direction is not possible. Moreover, by definition, uniaxial material has two equal eigenvalues of the dielectric constant tensor, but a biaxial material, such as $SmZ_A$ can have three distinct eigenvalues. While not definitive, we argue this is most straightforward explanation for the hysteresis observed in Figure 11. That is, cooling from the N phase results in a different layering structure (but the same director orientation) than heating from the $N_F$ phase. Ref. [3]presents evidence that upon cooling from N, the bookshelf geometry is favored over the parallel. While the data we present is consistent with a differing geometry depending on whether one cools from the N or heats from the $N_F$, it is more difficult to ascertain which is which. Lastly, during the cooling scan in Figure 11, about 4 degrees above the $N_F$ phase transition, a knee is observed (this is absent during heating). Below this knee $\boldsymbol{\epsilon}_{\perp}$ does not change with temperature (as is seen in the $N_F$ phase), so it seems logical that this is some pre-transitional effect, although its nature has not yet been determined.

Can the different layering geometries explain the hysteresis in the FT's? The magnetic case is more straightforward. If, below the transition (i.e. planar aligned director) the layers are bookshelf, the director distortion can commence without change in the layer normal. This is because in undistorted bookshelf the layer normal is the plane of the substrates. At very high field, the director becomes nearly homeotropic (i.e. perpendicular to the aligning surfaces). In this state, the layer normal will be degenerate in the plane of the aligning surfaces, i.e. not necessarily in the plane of the substrates. If this is the case, as the field is decreased, not only will the director be distorted, but also the layer normal. This additional distortion mode should increase the effective elastic constant leading to a larger critical field.

The preceding argument is more difficult to apply to the electric FT. This observation leads to a more significant indication. In the nematic phase, the electric and magnetic field induced FT's are essentially the same, other than a correction due to the electric susceptibility being so much larger than the magnetic. In the SmZ$_A$ phase, there is clear evidence that the two Freedericsksz transitions (electric and magnetic) are fundamentally, and qualitatively, different. What might explain such differing external field couplings? We argue the most likely cause by far is the anti-ferrolectric nature demonstrated for the SmZ$_A$ phase in DIO. [3]

The most obvious difference between the magnetic and electric field induced transitions in the SmZ$_A$ is the sharpness. In nematics, an ideal FT (infinite anchoring strength, zero pre-tilt) exhibits a discontinuity in slope. The transition shows rounding with either finite anchoring energy or pre-tilt (or both). How could these factors differentiate between magnetic and electric field? The antiferroelectric structure of SmZ$_A$ comprises a persistent polarization along the director and alternating periodically. In both the bookshelf and parallel geometries, a non-zero pre-tilt angle (at zero applied field) gives a small component of polarization perpendicular to the substrates. This component will have a linear coupling to an applied electric field, but not to any magnetic field. Thus, for any non-zero pretilt, we would expect a difference between the electric and magnetic field induced transitions in the SmZ$_A$ phase.

Above the FT threshold, there are intriguing implications of the dramatic increase in measured elastic constants in the SmZ$_A$ phase seen in Figure 5. As one enters the SmZ$_A$ phase from the N phase, $K_{11}$ and $K_{33}$ both become much larger (reaching more than 30X their value in the N phase), Yet, the increase is gradual, not abrupt; moreover, the increase begins above the phase transition temperature. This is in contrast to the well-known behavior whereby when approaching the N-SmA transition, the twist and bend elastic constants diverge, as these distortions are forbidden in the SmA state as they cannot be decoupled from layer compression. [6] Furthermore, in this case the divergence occurs in the higher temperature N phase; in the present case most of the increase is observed in the lower temperature SmZ$_A$ phase.

## Conclusions

When the cost of splay deformation becomes large, other manifestations of the Freedericksz transition may arise; in the most notable a periodic modulation arises in which twist deformation is substituted for splay. [2] Such modulations are expected when $K_{11}$ is more than twice $K_{22}$. Since DIO exhibits a large increase in $K_{11}$, the possibility of the periodic FT must be considered. Indeed, Figure 12 shows POM just above the magnetic field induced FT at 71 °C, where the splay constant is found to be 30 times its value in the N phase. There is clearly a modulated texture, although its periodicity appears to be poorly defined. A modulated texture is also observed above the electric field induced FT, although the character and spacing of the modulations appears to be different. In addition, the periodic FT requires splay much larger than twist. While the splay constant does increase dramatically, and there are as yet no measurements of the twist constant, it is argued [2] that the SmZ$_A$ layer structure precludes twist; it cannot be established that the periodic FT criteria are met in DIO. We lastly note a recent report of a periodic FT in the ferroelectric nematic state; while this work is in a related material, its origin is very different from what we report.

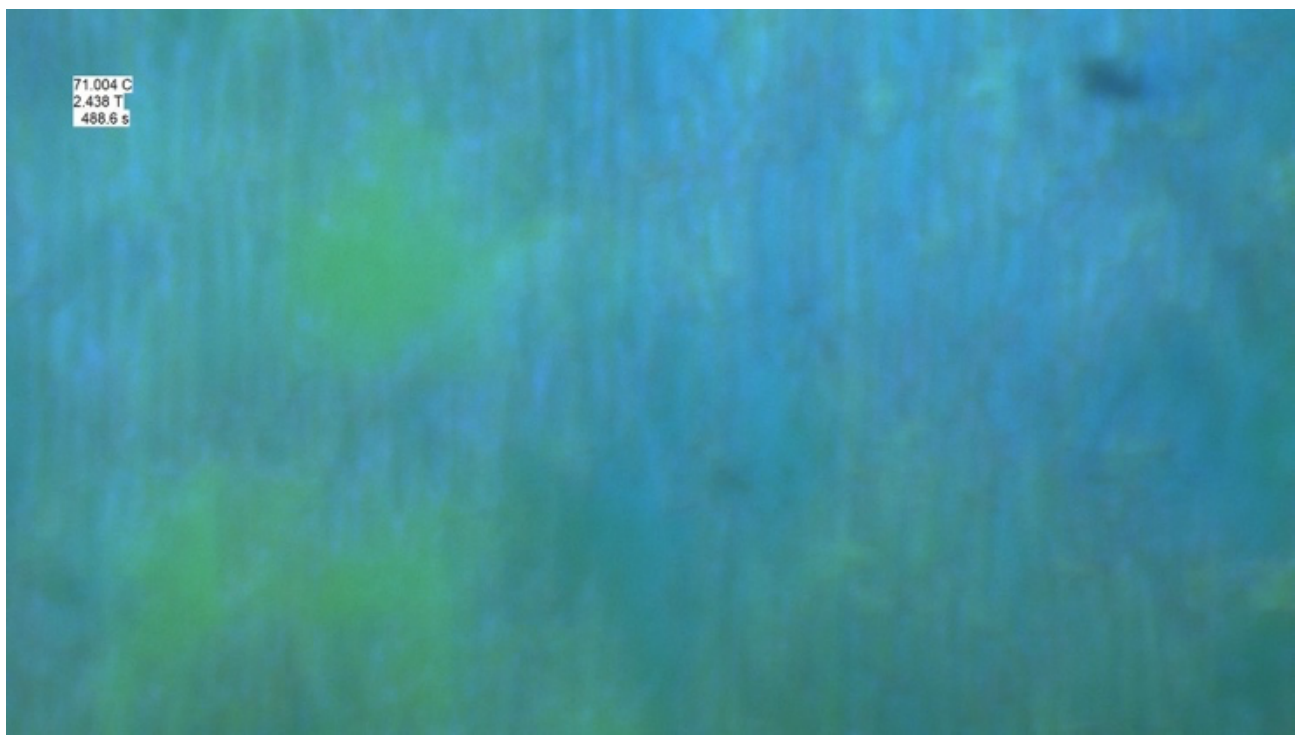

Figure 12 POM image of DIO in SmZa phase right above threshold for magnetic FT. Periodic striations perpendicular to rubbing direction are observed (not see in N phase).

It is somewhat surprising that the $SmZ_A$ phase exhibits field induced reorientation in a manner so similar to NLC materials. While measurements of the critical fields yield the elastic coefficients, in the absence of a comprehensive theory of the elasticity of the $SmZ_A$ one must perhaps be somewhat cautious in concluding precisely which elastic mode is being probed. Specifically, the coupling between director distortion (which also necessarily involves distortion of the polarization field) and the layer normal can be expected to play significant roles. Is this indicated by the almost thirty-fold increase in the cost of splay distortion in the $SmZ_A$ range? To address these, future planned work includes independent measurements via both the bend [26]and twist Freedericksz transitions. This will ascertain directly the anisotropy in the elastic constants, which should better illuminate the issue of the periodic FT. Moreover, in the undistorted state for the bend transition, the layer normal will be degenerate in the plane perpendicular to the field; how this affects the transition should be highly revealing.

## Conflicts of interest

There are no conflicts to declare.

## Acknowledgements

This work was supported by US National Science Foundation grant DMR-2210083.